\documentclass[aps,prl,twocolumn,floatfix,showpacs,superscriptaddress]{revtex4-1}
\usepackage{graphicx}
\usepackage{amssymb}
\usepackage{amsmath}
\usepackage{setspace}

\begin{document}

\title{Measurement of Rb 5P$_{3/2}$ scalar and tensor polarizabilities in a 1064~nm light field}
\author{Yun-Jhih~Chen}
\affiliation{Department of Physics, University of Michigan, Ann Arbor, MI 48109}
\author{Lu\'is~Felipe~Gon\c{c}alves}
\affiliation{Department of Physics, University of Michigan, Ann Arbor, MI 48109}
\affiliation{Instituto de F\'isica de S\~ao Carlos, Universidade de S\~ao Paulo, CP 369, 13560-970, S\~ao Carlos, SP, Brasil}
\author{Georg~Raithel}
\affiliation{Department of Physics, University of Michigan, Ann Arbor, MI 48109}
\date{\today}

\begin{abstract}
We employ doubly-resonant two-photon excitation into the 74S Rydberg state to spectroscopically measure the dynamic scalar polarizability, $\alpha_0$, and tensor polarizability, $\alpha_2$, of rubidium 5P$_{3/2}$. To reach the necessary high intensities, we employ a cavity-generated 1064~nm optical-lattice light field, allowing us to obtain intensities near $2 \times10^{11}$~W/m$^2$. In the evaluation of the data we use a self-referencing method that renders the polarizability measurement largely free from the intensity calibration of the laser light field. We obtain experimental values $\alpha_0=-1149~(\pm2.5\%)$ and $\alpha_2=563~(\pm4.2\%)$, in atomic units. Methods and results are supported by simulations.
\end{abstract}
\maketitle

Polarizabilities of atomic energy levels govern the response of an atom to an external electric field and are a fundamental property to be considered in atom-trapping and precision-measurement experiments. For example, the polarizabilities are required for determining magic wavelengths of state-insensitive trapping in optical lattices \cite{Arora.2007,Goldschmidt.2015}. Theoretical calculations of dynamic polarizabilities are complicated, and yet available experimental measurements might carry large uncertainties due to the difficulty in calibrating the field strength experienced by the atoms. Here, we report a measurement of rubidium scalar and tensor polarizabilities conducted in a strong 1064~nm light field, where the magnetic sublevels of Rb 5P$_{3/2}$ are resolved and the data analysis is largely free from the calibration of laser intensity. Despite the fact that 1064~nm optical traps for rubidium atoms are widely used in cold-atom trapping, to our best knowledge there is no other such experimental measurement. Our work is not only applicable to experiments utilizing Rb 5P$_{3/2}$ levels in 1064~nm laser traps, but also serves as an experimental test for validating and improving existing theoretical models for the polarizability.

\begin{figure}[!b]
\vspace{-12pt}
\begin{centering}
\includegraphics[width=3.4in]{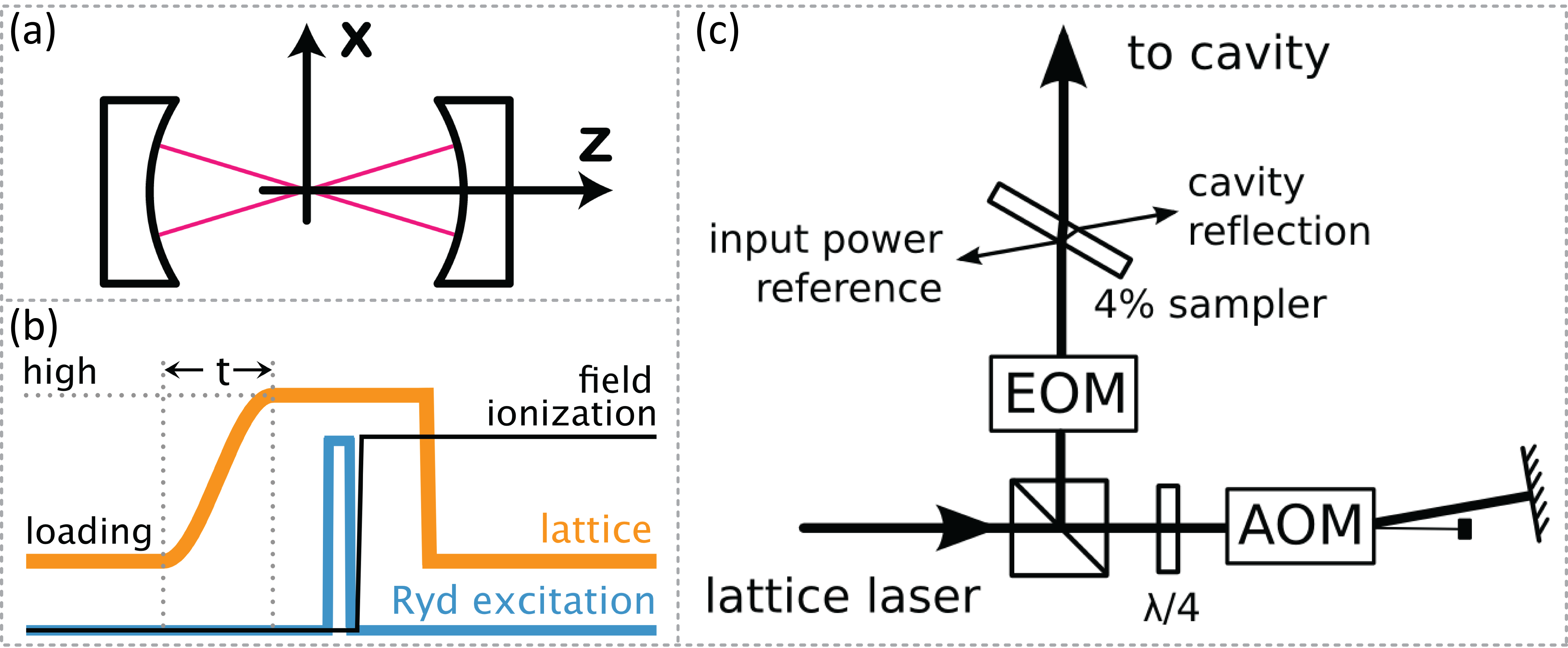}
\vspace{-12pt}
\caption{(color online). Experimental details. (a) A near-concentric cavity, which has a focus at its cavity center, is used to generate a GHz-deep 1064~nm optical-lattice laser trap. (b) Timing sequence showing lattice intensity, Rydberg-atom excitation and field ionization pulses vs time ($\rm t=30\mu\rm s$, repetition rate = 100~Hz). (c) Key components of the PDH cavity stabilization. }\label{fig1}
\end{centering}
\vspace{-12pt}
\end{figure}

In the presence of an optical field, an atom is polarized and its energy levels are shifted. The applicable atom-field interaction Hamiltonian, $\hat{H}_{\rm E}$, in a linearly-polarized electric field with amplitude $E_0$ is
\begin{equation}
-\frac{E^2_0}{4} \sum_{m_J} \vert J, m_J \rangle \langle J, m_J \vert \Big[\alpha_0(\omega)+\alpha_2(\omega)\frac{3m_J^2-J(J+1)}{J(2J-1)}\Big]\label{acstark}
\end{equation}
where $\alpha_0(\omega)$ and $\alpha_2(\omega)$ are frequency-dependent a.c. scalar and tensor polarizabilities, and $J$ and $m_J$ are quantum numbers of the total electronic angular momentum. The dynamic polarizability of the rubidium ground-state 5S$_{1/2}$ depends only on $\alpha_0$ ($\alpha_2$ is zero because $J=\frac{1}{2}$). For 5P$_{3/2}$, both $\alpha_0$ and $\alpha_2$ contribute to the polarizability. The full Hamiltonian includes the hyperfine structure, $\hat{H}=\hat{H}_{\rm{HFS}}+\hat{H}_{\rm E}$, with
\begin{eqnarray}\nonumber
\hat{H}_{\rm{HFS}}&&=A_{\rm{HFS}}~\hat{\textbf{I}}\cdot\hat{\textbf{J}}+\\
&&B_{\rm{HFS}}~\frac{3(\hat{\textbf{I}}\cdot\hat{\textbf{J}})^2+\frac{3}{2}\hat{\textbf{I}}\cdot\hat{\textbf{J}}-I(I+1)J(J+1)}{2IJ(2I-1)(2J-1)}\label{abhfs}
\end{eqnarray}
where $\hat{\textbf{I}}$ is the nuclear spin, $A_{\rm{HFS}}$ is the magnetic-dipole and $B_{\rm{HFS}}$ the electric-quadrupole hyperfine constant. $A_{\rm{HFS}}$ and $B_{\rm{HFS}}$ are well known for $^{87}$Rb and $^{85}$Rb \cite{Steck.2013,Steck.2010}. The octupole contribution is omitted because it is too small to be observed here.

We measure $\alpha_0$ and $\alpha_2$ using Rydberg two-photon excitation spectroscopy. The data analysis is based on linear fits of spectral data sets in a modified a.c. Stark map, in which the frequencies of the two excitation lasers are plotted against each other, with the unknown polarizabilities as fitting parameters. Our method has the strength that it does not require a precise calibration of the 1064~nm laser intensity at the atom trapping site. Also, light shifts are in the range of several GHz, which is important for a precise measurement of $\alpha_0$ and $\alpha_2$ of Rb 5P$_{3/2}$ (level width 6~MHz).

We utilize an in-vacuum, moderate-finesse ($\approx600$), near-concentric optical cavity [see Fig.~1~(a)] at 1064~nm to generate deep optical-lattice potentials in a linearly-polarized light field~(for details see Ref.~\cite{Chen.2014}). We load the lattice at low intensity directly from a $^{87}$Rb magneto-optical trap (MOT), and then increase the lattice power after the MOT light is pulsed off [timing see
Fig.~\ref{fig1}~(b)]. The smooth sinusoidal lattice-intensity ramp compresses the atom distribution in the lattice wells, where light shifts reach several GHz. Atoms are excited into Rydberg states while the lattice is held at a high, constant intensity. Rydberg atoms are detected by field ionization~\cite{RydbergAtoms} and ion counting. After detection, the lattice is switched back to the loading intensity for the next experimental cycle. During the loading phase, the 5S$_{1/2}$ ground-state atoms have an estimated trap frequency of about 700~kHz along $Z$ and about 7~kHz along $X$ and $Y$. Hence, the 30~$\mu$s lattice ramp leads to adiabatic compression in the $Z$ and mixed adiabatic/diabatic compression in the transverse directions.

\begin{figure}[!b]
\vspace{-12pt}
\begin{centering}
\includegraphics[width=3.4in]{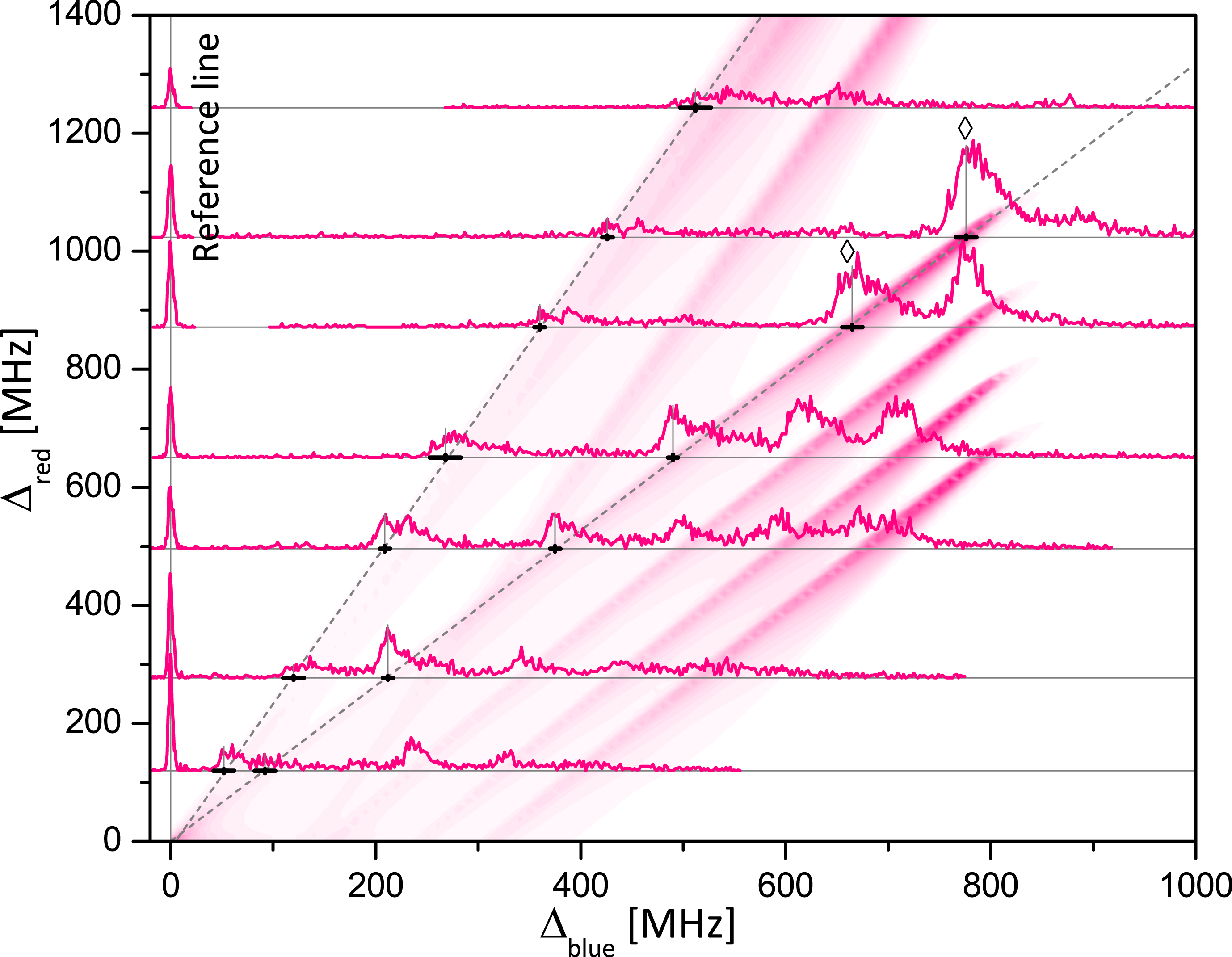}
\vspace{-12pt}
\caption{(color online). Experimental spectra overlaid over a simulated spectral-density plot.  The grey drop lines mark the $(\Delta_{\rm{blue}}, \Delta_{\rm{red}})$ coordinates of the peaks of the evaluated spectral lines.  The bold, short black lines underneath the evaluated spectral lines indicate the uncertainties of the peak coordinates.  The dashed grey lines show linear fits through the marked positions $(\Delta_{\rm{blue}}, \Delta_{\rm{red}})$. For the simulated density plot, the peak lattice intensity is 1.8$\times 10^{11}$W/cm$^2$, $\alpha_0(5\rm P_{3/2})=-1149.3$, $\alpha_2(5\rm P_{3/2})=563.3$, and $\alpha_0(5\rm S_{1/2})=687.3$ (polarizabilities in atomic units).  }\label{fig2}
\end{centering}
\vspace{-12pt}
\end{figure}

The in-vacuum near-concentric lattice resonator is stabilized to the 1064~nm trap laser (short-term bandwidth 100~kHz) by the Pound-Drever-Hall (PDH) scheme~\cite{Black.2001} [see Fig.~\ref{fig1} (c)]. The electro-optic modulator (EOM) generates PDH frequency side bands. The reflection from the cavity is sampled by a beam sampler for synthesizing the PDH error signal. The PDH feedback circuit has two outputs: a high-voltage slow feedback is applied to a piezo, which translates one of the cavity mirrors, and a fast feedback frequency-modulates the acousto-optic modulator (AOM), which compensates the rapid frequency fluctuations of the trap laser. The lattice intensity ramp is generated by amplitude-modulating the AOM. To ensure acceptable performance of the lattice cavity lock during lattice compression, the amplitude of the PDH error signal is normalized in real time by a trap-laser power reference. The normalization circuit is composed of op-amp logarithmic and exponential amplifiers. Our method to stabilize the optical cavity at high intra-cavity power differs from the strong-weak beam method~\cite{Edmunds.2013} and from methods that use a separate reference cavity~\cite{Hamilton.2015}.

The polarizabilities are derived from two-photon stepwise (resonant) excitation signals from the light-shifted 5S$_{1/2}$ ground state through a light-shifted 5P$_{3/2}$ sublevel into the 74S$_{1/2}$ Rydberg state. For each of the selected lattice intensities, the two-photon excitation spectra are taken for a set of fixed lower-transition detunings, $\Delta_{\rm{red}}$, by recording Rydberg counts as a function of the upper-transition detuning, $\Delta_{\rm{blue}}$. We reference $\Delta_{\rm{blue}}$ to a narrow spectral line that corresponds to off-resonant two-photon excitation of Rydberg atoms outside the lattice. This line serves as a well-defined reference point. For each lattice intensity, the spectra are arranged in waterfall plots in which the $y$-displacement is given by $\Delta_{\rm{red}}$, while $\Delta_{\rm{blue}}$ is plotted along the $x$-axis. In Fig.~\ref{fig2} we show such a modified a.c. Stark map measured for a transmitted lattice power of 20~mW, which corresponds to a peak intracavity intensity at the lattice sites of about 1.8$\times10^{11}$~W/m$^2$. The uncertainty of the transmitted lattice power is about 8\% and that of the peak intracavity intensity is even larger. It is an essential advantage of our method that these uncertainties do not affect the values of $\alpha_0$ and $\alpha_2$ obtained from the data. This is because $\alpha_0$ and $\alpha_2$ are derived from slopes in the modified a.c. Stark map, not from absolute level positions. In the following, we explain this self-referencing method in detail.

\begin{figure}[!b]
\vspace{-12pt}
\begin{centering}
\includegraphics[width=3.4in]{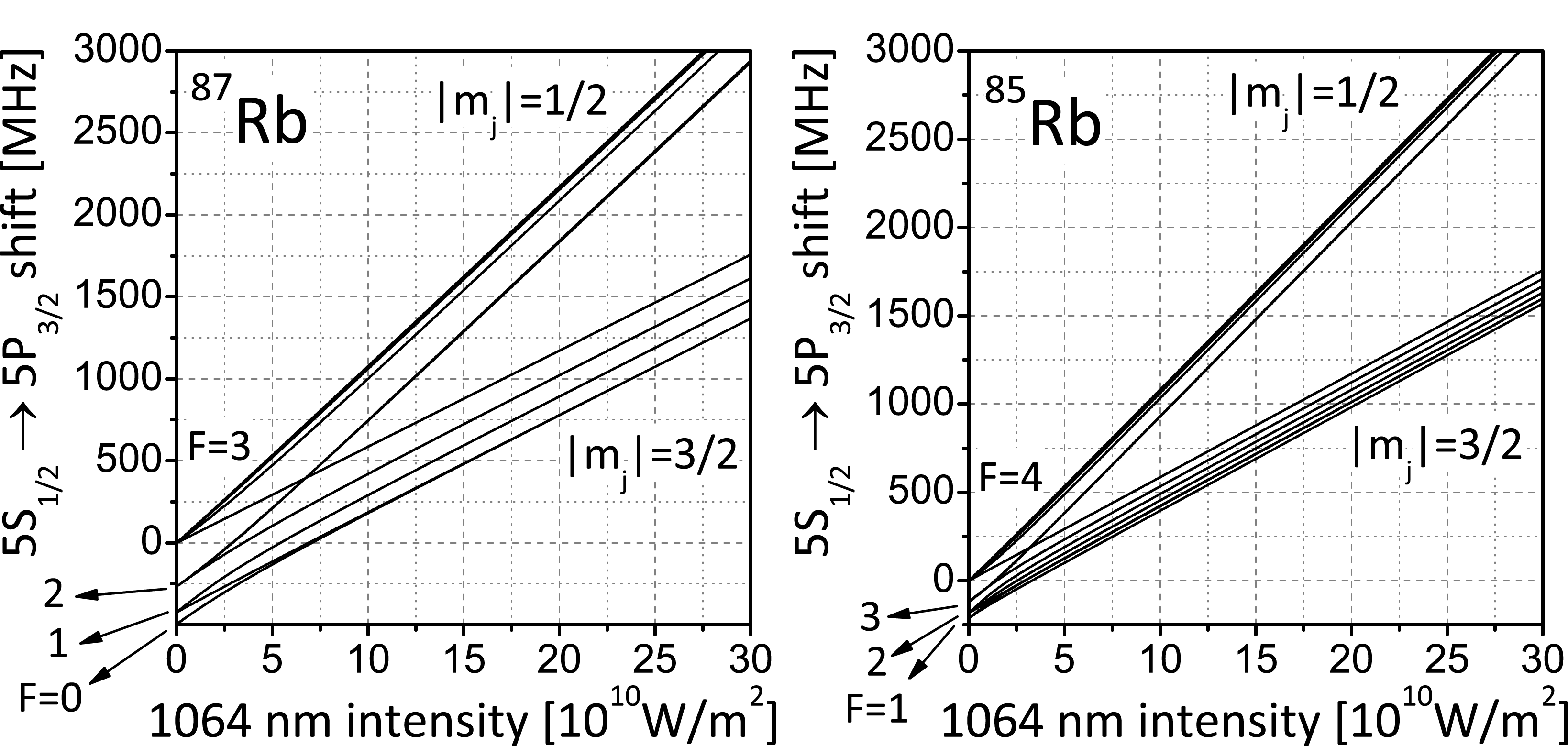}
\vspace{-12pt}
\caption{A.c. Stark shifts of the transitions $5\rm S_{1/2}\to 5\rm P_{3/2}$ of $^{87}$Rb ($^{85}$Rb) relative to the transition $5\rm S_{1/2},~F=2(3)\to 5\rm P_{3/2},~F'=3(4)$, calculated for $\alpha_0(5\rm S_{1/2})=700$, $\alpha_0(5\rm P_{3/2})=-1100$, and $\alpha_2(5\rm P_{3/2})=550$ atomic units. }\label{fig3}
\end{centering}
\vspace{-12pt}
\end{figure}

In Fig.~\ref{fig3} we show the light shifts of the 5S$_{1/2}\to$5P$_{3/2}$ transitions versus intensity, $I_0=\frac{1}{2}c\epsilon_0E_0^2$, calculated by diagonalization of the full Hamiltonian (Eqs.\ref{acstark} and~\ref{abhfs}) using approximate values of $\alpha_0$ and $\alpha_2$. The $\alpha_0$ and $\alpha_2$ are the same for both rubidium isotopes, because they only depend on electron-field coupling. Since the 5S$_{1/2}$ state has no a.c.-split sublevels, the splitting is only due to the 5P$_{3/2}$ state, which has 16 sublevels for $^{87}$Rb and 24 for $^{85}$Rb. The plots exhibit three intensity regimes. In the weak-field regime, $\hat{H}_{\rm{HFS}}$ dominates, and $|F,m_F\rangle$ ($\hat{\textbf{F}}=\hat{\textbf{J}}+\hat{\textbf{I}}$) is the ``good'' basis. Levels with the same $F$ and $|m_F|$-values are degenerate, and their energy shifts are linear in intensity. At intermediate intensity neither $\hat{H}_{\rm{HFS}}$ or $\hat{H}_{\rm E}$ dominates, and the energy levels are generally nonlinear. The level crossing at $7 \times 10^{10}$~W/m$^2$ (for $^{87}$Rb) depends mostly on $\alpha_2$ and the zero-field hyperfine splittings. For the experimental determination of $\alpha_0$ and $\alpha_2$ we utilize the strong-field regime, which is analogous with the ``Paschen-Back'' regime of the Zeeman effect. With our near-concentric cavity setup, we comfortably reach the strong-field regime.

In the strong-field regime $\hat{H}_{\rm{E}}$ dominates, and $|I J m_I m_J \rangle$ becomes the ``good'' basis. The energy levels become linear functions of intensity again, and they separate into two groups of fixed $|m_J|$. Since $\alpha_2>0$, the group with higher energy includes levels with $|m_J|=\frac{1}{2}$, while the group with lower energy has $|m_J|=\frac{3}{2}$. Within each subgroup, the energies are split by the residual hyperfine perturbation. For example, for $^{87}$Rb ($I=\frac{3}{2}$ and $J=\frac{3}{2}$), in the subspace $|m_J|=\frac{3}{2}$ the off-diagonal terms of $\hat{\textbf{I}}\cdot\hat{\textbf{J}}$ and $(\hat{\textbf{I}}\cdot\hat{\textbf{J}})^2$ vanish in the $|m_Im_J\rangle$ basis, and the residual hyperfine-induced shifts are given by the diagonal terms (which follow from $m_I m_J = \pm\frac{9}{4}$ or $\pm\frac{3}{4}$). Thus, the lower subgroups in Fig.~\ref{fig3} appear nearly equally spaced, due to the leading magnetic-dipole hyperfine term $A_{\rm{HFS}}~\hat{\textbf{I}}\cdot\hat{\textbf{J}}$. The deviation from an equal spacing is due to the electric-quadrupole hyperfine term, whose strength is several percent of that of the magnetic-dipole term. In the subgroups with $|m_J|=\frac{1}{2}$ the off-diagonal elements of $\hat{\textbf{I}}\cdot\hat{\textbf{J}}$ are generally non-zero, and the residual hyperfine shifts are not approximately equidistant. In our experiment we choose $^{87}$Rb, because the number of $m_I$ sublevels is less and the residual hyperfine splittings in the Paschen-Back regime are larger than for $^{85}$Rb. However, the method also applies to $^{85}$Rb.

\begin{figure}[!b]
\vspace{-12pt}
\begin{centering}
\includegraphics[width=3.4in]{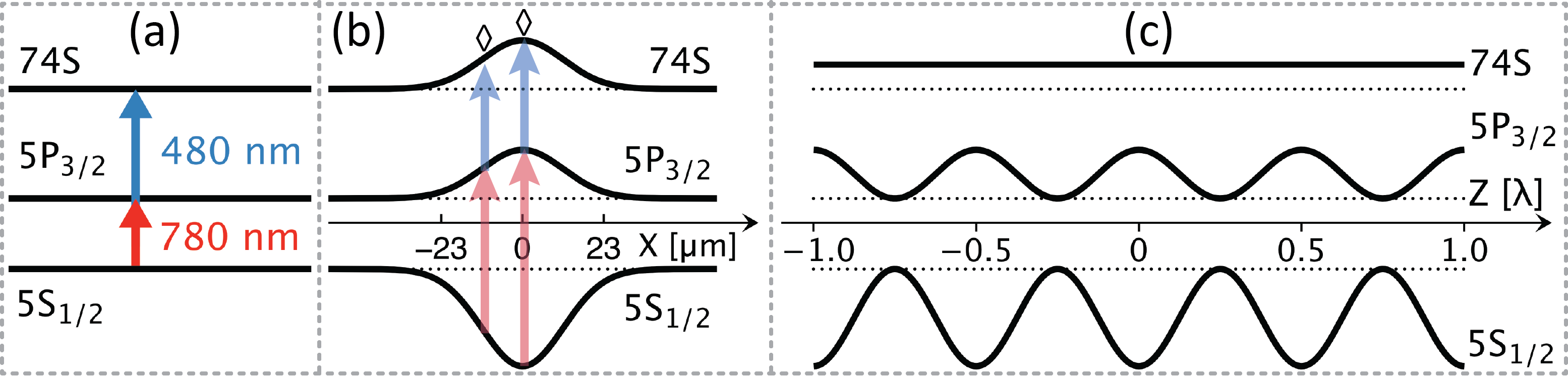}
\vspace{-12pt}
\caption{(color online). Relevant levels of the two-photon double-resonant Rydberg excitation.(a) Outside lattice. (b,c) Light-shifted levels (solid lines) in transverse (b) and $Z$-direction (c) in the 1064~nm lattice. The ``magic'' 74S state has a constant light shift along $Z$. Dotted lines indicate the unshifted levels. The intervals between levels are not to scale. The arrows illustrate several instances of doubly-resonant excitation on one of the lines in Fig.~\ref{fig2} (see $\diamondsuit$ in Fig.~2).}
\label{fig4}
\end{centering}
\vspace{-12pt}
\end{figure}

The 5S$_{1/2}$ and 5P$_{3/2}$ levels exhibit a local response to the 1064~nm lattice light field, {\sl{i.e.}} their light shifts are given by
$-(\alpha/4) |E_0(\boldsymbol{R})|^2$,
where $E_0(\boldsymbol{R})$ is the field amplitude at the center-of-mass location $\boldsymbol{R}$ of the atom. Due to their size, Rydberg atoms have a non-local response to the field. The Rydberg-atom light shift is
\begin{equation}
V_{\rm{ad}}(\boldsymbol{\rm R})=-\frac{1}{4}\alpha_e\int |E_0(\boldsymbol{\rm{R+r}})|^2|\psi(\boldsymbol{\rm r})|^2d^3r
\label{vad1}
\end{equation}
where $\alpha_e=-545$~at.~un. follows from the free-electron ponderomotive energy \cite{Dutta.2000}. Thus, the Rydberg-atom light shift is an average of the free-electron response, with the Rydberg electron's probability density $|\psi(\boldsymbol{\rm r})|^2$ as a weighting factor. While generally $\psi(\boldsymbol{\rm r})$ also depends on ${\bf {R}}$, for the non-degenerate $n$S Rydberg levels it is $\psi(\boldsymbol{\rm r}) = \langle {\bf {r}} \vert n $S$ \rangle$. The averaging in $z$-direction is important, because the size of the Rydberg atom is on the order of the lattice period. The averaging in $\rho$ direction is not important, because the cavity-mode waist $w_0$ is much greater than the size of the Rydberg atom. For a few ``magic'' states, such as Rb~74S$_{1/2}$ in a one-dimensional 1064~nm lattice, $V_{\rm{ad}}$ only depends on $\rho$ and not on $Z$. For the magic states,
\begin{equation}
V_{\rm{ad}}(\rho,Z) = V_{\rm{ad}}(\rho) = -\frac{1}{4}\frac{\alpha_e}{2}E^2_{\rm{ max}}\exp(\frac{-2\rho^2}{w_0^2})
\label{vad2}
\end{equation}
where $E_{\rm{max}}$ is the peak field amplitude in the entire lattice. In Fig.~\ref{fig4} we show the position dependence of the relevant light shifts as a function of atomic center-of-mass coordinates $X$ and $Z$. The peaks in Fig.~\ref{fig2} correspond to doubly-resonant excitation of 74S$_{1/2}$ through one of the multiple 5P$_{3/2}$ sublevels.

\begin{figure}[!t]
\vspace{-12pt}
\begin{centering}
\includegraphics[width=3.4in]{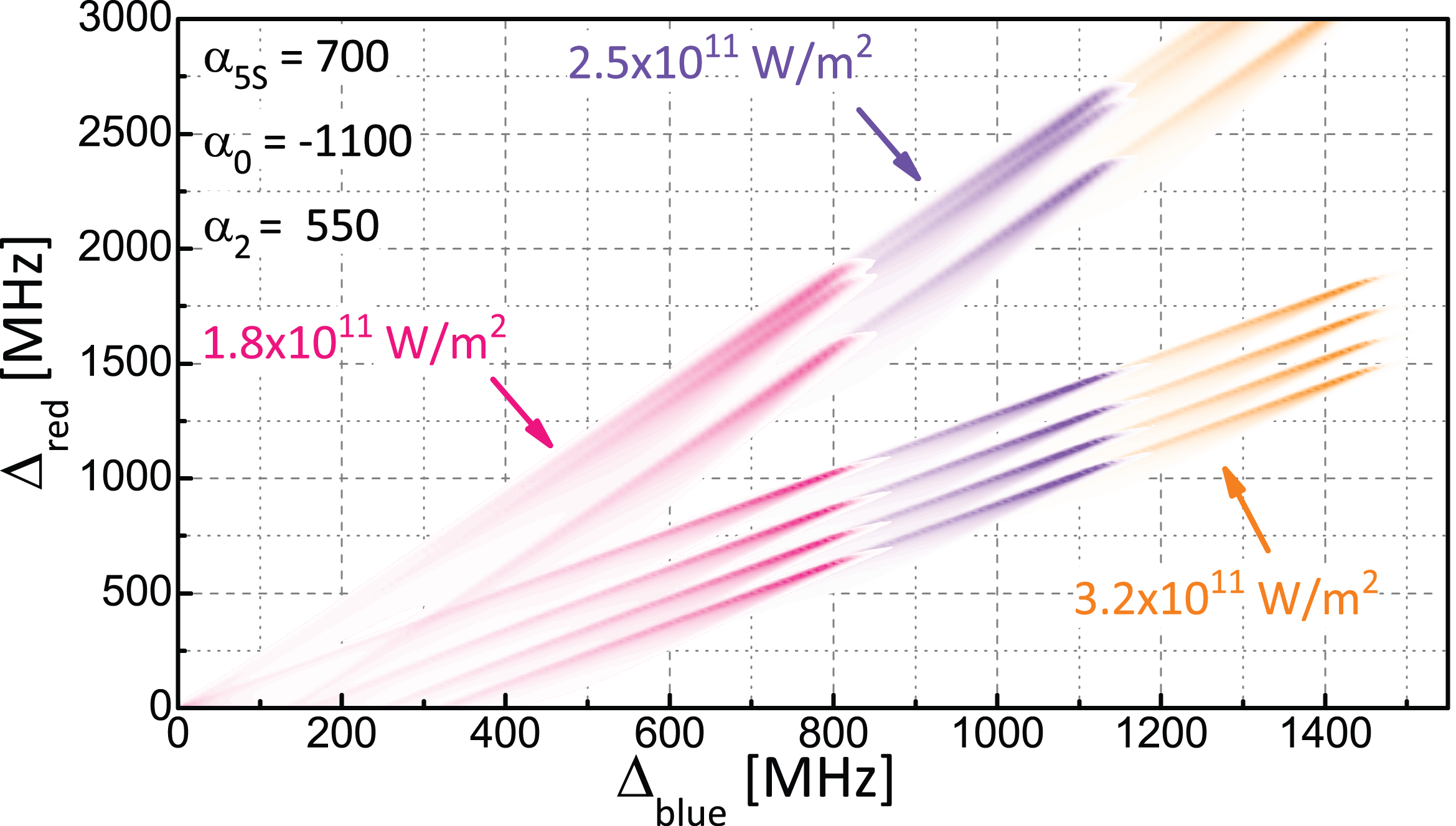}
\vspace{-12pt}
\caption{(color online). Simulated spectra. We overlay three calculations for the indicated lattice intensities.}\label{fig5}
\end{centering}
\vspace{-12pt}
\end{figure}

The density plot in the background in Fig.~\ref{fig2} shows the result of a simulation. The simulation accounts for the center-of-mass thermal distribution of the atoms in the optical lattice (temperature $T$), which causes most of the spectral line broadening. Doppler shifts are negligible and are ignored. The simulated count rate is an integral over the lattice volume that includes the Boltzmann factor, $\exp[-\alpha_{\mathrm{5S}}(E_{\rm{max}}^2-E_0^2(\rho,Z))/(4 k_{\rm B} T)]$, and the lower-transition saturation parameter, $0.5 s /(1+ s + 4(\frac{\Delta}{6~{\rm {MHz}} })^2)$. There, $s=I/I_{\rm{sat}}$ with saturation intensity $I_{\rm {sat}}$ and position-dependent intensity $I$. Also, $\Delta$ is the detuning of the 780~nm laser from the light-shifted, position-dependent 5S$_{1/2}\to$5P$_{3/2}$ transition frequency. The simulated count rate is summed over all intermediate 5P$_{3/2}$ states. Figure~\ref{fig5} shows a compilation of three simulations for $^{87}$Rb for different peak lattice intensities and the same polarizabilities. The slopes in the modified a.c. Stark map are insensitive to the peak lattice intensity. The peak lattice intensity affects the signal strength and the $\Delta_{\rm{blue}}$ cutoff, where the doubly-resonant excitation condition is met at the field maxima (which are at locations $\rho=0$ and $Z=k \times 532$~nm with integer $k$) for all 5P$_{3/2}$ sublevels. The peak lattice intensity in the experiment can be estimated from the $\Delta_{\rm{blue}}$ cutoff; this intensity is, however, not required to extract the polarizability values.

As can be seen from experimental and simulated spectra, the lattice-shifted spectral lines are asymmetric and triangular. The line shape results from the variation of the Rydberg light shift along surfaces of fixed intensity; the variation is due to the non-local response following Eqs.~\ref{vad1} and~\ref{vad2}. The line shape is discussed in~\cite{supp}.
We use the positions ($\Delta_{\rm{blue}},\Delta_{\rm{red}}$) of the peaks of the spectral lines as primary markers, as shown in Fig.~\ref{fig2}. These peaks correspond to combinations of intensity $I$, $\Delta_{\rm{blue}}$ and $\Delta_{\rm{red}}$ for which the double-resonance condition for excitating 74S$_{1/2}$ is met at locations $Z=k \times 532$~nm with integer $k$~\cite{supp}.

In the modified a.c. Stark map, the peak positions ($\Delta_{\rm{blue}},\Delta_{\rm{red}}$) satisfy
\begin{eqnarray}
y&=&\Delta_{\rm{red}}=\Delta_{\rm{5S5P}}=\frac{1}{4}(\alpha_{\rm{5S}}-\alpha_{\rm{5P}})E_0^2(\rho,Z=0) \nonumber \\
x&=&\Delta_{\rm{blue}}=\Delta_{\rm{74S}}+\Delta_{\rm{5S5P}}-\Delta_{\mathrm{5P}}\nonumber \\
&=&\Delta_{\rm{5S5P}}+\frac{1}{4}(\alpha_{\rm{5P}}-\alpha_{\rm{74S}})E_0^2(\rho,Z=0)=
y~\frac{\alpha_{\rm{5S}}-\alpha_{\rm{74S}}}{\alpha_{\rm{5S}}-\alpha_{\rm{5P}}} \nonumber
\end{eqnarray}
where $E_0(\rho,Z=0)$ is the maximal lattice field at a distance $\rho$ from the axis.  The slopes of the levels are independent of $E_0$ and are
\begin{equation}
\frac{{\rm d} y}{{\rm d} x}=\frac{\alpha_{\rm{5S}}-\alpha_{\rm{5P}}}{\alpha_{\rm{5S}}-\alpha_{\rm{74S}}}
\end{equation}
In the high-field (Paschen-Back) regime, the $\alpha_{\mathrm{5P}}$ of the highest energy levels in each subgroup of $|m_j|=\frac{1}{2}$ and $|m_j|=\frac{3}{2}$ are, in terms $\alpha_0$ and $\alpha_2$,
\begin{eqnarray}
\alpha_{\mathrm{5P}}(|m_j|=\frac{1}{2})
&=&\alpha_0-\alpha_2\\
\alpha_{\mathrm{5P}}(|m_j|=\frac{3}{2})
&=&\alpha_0+\alpha_2
\end{eqnarray}
The differential and average slopes are
\begin{eqnarray}
\frac{\mathrm d y}{\mathrm d x}\Bigg|_{|m_j|=\frac{1}{2}}-\frac{\mathrm d y}{\mathrm d x}\Bigg|_{|m_j|=\frac{3}{2}}&=&\frac{2\alpha_2}{\alpha_{\mathrm{5S}}-\alpha_{\mathrm{74S}}}\label{slopediff}\\
\frac{1}{2}\Bigg(\frac{\mathrm d y}{\mathrm d x}\Bigg|_{|m_j|=\frac{1}{2}}+\frac{\mathrm d y}{\mathrm d x}\Bigg|_{|m_j|=\frac{3}{2}}\Bigg)&=&\frac{\alpha_{\mathrm{5S}}-\alpha_0}{\alpha_{\mathrm{5S}}-\alpha_{\mathrm{74S}}}\label{slopeavg}
\end{eqnarray}

\begin{table}[!b]
\vspace{-12pt}
	\centering
	\caption{Experimental 5P$_{3/2}$ scalar and tensor polarizabilities in 1064~nm light field. The uncertainties are reading and fitting uncertainties. Polarizabilities are in atomic units.}
		\begin{tabular}{lll}
		 \hline
		  parameter& value \quad\quad& uncertainty \\ \hline
      Experimental slope $|m_j|=\frac{3}{2}$ \quad & 1.33 & 4.19$\times 10^{-3}$ \\
      Experimental slope $|m_j|=\frac{1}{2}$ \quad & 2.50 & 4.47$\times 10^{-2}$ \\
      differential slope & 1.17 & 4.49$\times 10^{-2}$\\
      average slope & 1.91 & 2.24$\times 10^{-2}$\\
      $\alpha_0$ (experiment) & -1149 & 22\\
      $\alpha_2$ (experiment)& 563 & 22\\
       \hline \hline
		\end{tabular}
	\label{tab1}
\end{table}

The slopes $dy/dx$ are determined by linear fitting of the sets of peaks ($\Delta_{\rm{blue}},\Delta_{\rm{red}}$) associated with the respective lines in the modified a.c. Stark map. For $|m_j|=\frac{3}{2}$ we force the $y$-intercept to zero because this level has a fixed slope through all field regimes and passes through the origin. For $|m_j|=\frac{1}{2}$ we fit both the slope and the intercept. This is because in the weak field regime the $|m_j|=\frac{1}{2}$ level connects to $|F=3,m_F=0\rangle$, which has $\alpha_{\rm{5P}}=\alpha_0-\frac{4}{5}\alpha_2$. Hence, the high-field fit of the $|m_j|=\frac{1}{2}$ level yields a slight negative $y$-intercept, which is on the order of $-30$~MHz for a peak lattice intensity of $1.8\times10^{11}$W/m$^2$. We fit the slopes for measurements at several peak lattice intensities and excitation-pulse durations~\cite{supp}. The numbers of slope samples are 5 and 4 for $|m_j|=\frac{3}{2}$ and $|m_j|=\frac{1}{2}$, respectively. The weighted average slopes are summarized in Table \ref{tab1}. The $\alpha_{\rm{5S}}$ and $\alpha_{\rm{74S}}$ are, for the present purpose, precisely known ($\alpha_{\rm{5S}} = 687.3(5)$ \cite{Arora.2012} and $\alpha_{\rm{74S}} = -272.5(5)$ \cite{Dutta.2000}, in atomic units). Eqs.~\ref{slopediff} and \ref{slopeavg} then yield $\alpha_0=-1149$ and $\alpha_2=563$~at.~un.

The uncertainties listed in Table~\ref{tab1} only reflect the linear-fitting uncertainty and the reading uncertainties associated with the  determination of the peak markers ($\Delta_{\rm{blue}},\Delta_{\rm{red}}$). The calibration uncertainties for $\Delta_{\rm{blue}}$ and $\Delta_{\rm{red}}$ are 1.6\% and 0.2\%, respectively. Adding all relative uncertainties in quadrature, the final uncertainties are 2.5\% for $\alpha_0$ and 4.2\% for $\alpha_2$~\cite{supp}. Our result is in good agreement with theoretical values $\alpha_0=-1111.62$ and $\alpha_2=557.31$ obtained in Ref.~\cite{Arora.2012}. Also, we have adopted the theoretical calculation of $\alpha_{5\rm S} = 687.3(5)$ in~\cite{Arora.2012}, instead of an earlier calculation from~\cite{Marinescu.1994} and an experimental value $\alpha_{5\rm S} = 769(61)$~\cite{Bonin.1993}.

Our method could be adapted to other atoms in deep optical lattices and at different wavelengths, such as for Cs in a deep 1064~nm lattice or Rb in a high-power CO$_2$-laser lattice. Instead of using a magic Rydberg state, one may utilize low-lying Rydberg states that are sufficiently small that they exhibit an approximately local response to the field (given by the free-electron polarizability). A different approach to polarizability measurement in 1064~nm laser fields can be found in \cite{Edmunds.2014}.

The experimental values for the scalar and tensor polarizabilities of 5P$_{3/2}$ are immediately useful in experiments that require on-resonant transitions through 5P$_{3/2}$, such as two-photon preparation of lattice-mixed hydrogenic Rydberg states in deep 1064~nm lattices \cite{Younge.2009}, Rydberg-EIT in 1064~nm optical traps, and evaporative cooling utilizing 1064~nm light fields.

\begin{acknowledgments}
We thank Dr. B.~K.~Sahoo for providing theoretical values of scalar and tensor polarizabilities. LFG acknowledges S\~ao Paulo Research Foundation (FAPESP 2014/09369-0). This work is supported by NSF Grant No. PHY-1205559.
\end{acknowledgments}

\end{document}


\title{Supplementary material: Measurement of Rb 5P$_{3/2}$ scalar and tensor polarizabilities in a 1064~nm light field}
\author{Yun-Jhih~Chen}
\affiliation{Department of Physics, University of Michigan, Ann Arbor, MI 48109}
\author{Lu\'is~Felipe~Gon\c{c}alves}
\affiliation{Department of Physics, University of Michigan, Ann Arbor, MI 48109}
\affiliation{Instituto de F\'isica de S\~ao Carlos, Universidade de S\~ao Paulo, CP 369, 13560-970, S\~ao Carlos, SP, Brasil}
\author{Georg~Raithel}
\affiliation{Department of Physics, University of Michigan, Ann Arbor, MI 48109}
\date{\today }
\maketitle

\section{Spectral line shape}\label{line shape}
The triangular shape of the lattice spectral lines results from the non-locality of the Rydberg atom's response to the field, which leads to a variation of the Rydberg-level light shift along surfaces of constant intensity within the optical lattice. This is explained in the following.

\begin{figure*}
\begin{centering}
\includegraphics[width=5in]{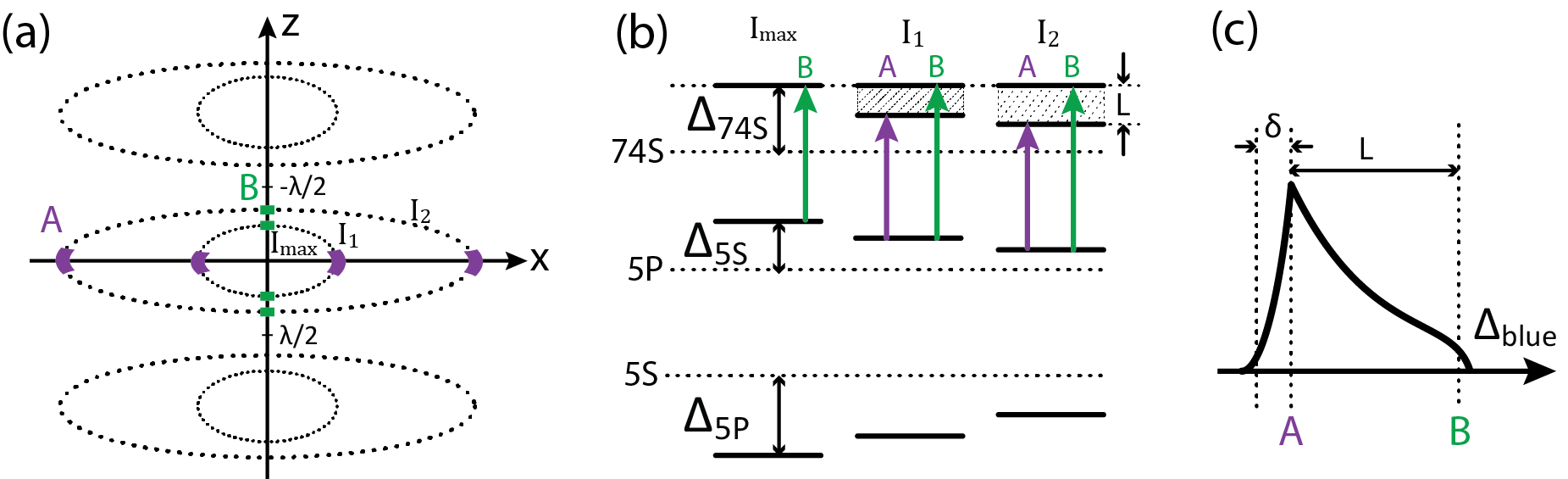}
\caption{(color online). (a) Constant-intensity ``ellipses'' for intensities $I_{\rm{max}}>I_1>I_2$. The eccentricity of the ellipses increases with decreasing intensity. (b) Corresponding light shifts of the relevant levels. The shift of the Rydberg state varies over a range $L$ that increases with decreasing intensity. (c) Typical triangular line shape. The left width $\delta$ is small and given by the level width of the intermediate state (6~MHz) and the laser line widths ($\sim 1$~MHz). The right width is largely given by $L$.}
\label{figs1}
\end{centering}
\end{figure*}

Along the lattice axis ($Z$-axis) the lattice laser intensity sinusoidally varies with a period of $\frac{\lambda}{2}=$532~nm. In our coordinate system, the lattice-intensity maxima are located at $Z=k \frac{\lambda}{2}$, with integer $k$. In the transverse direction, i.e., the $X$- and $Y$-axes, the lattice intensity follows a Gaussian profile $I(\rho,0)=I_{\rm{max}}\exp(\frac{-2 \rho^2}{w_0^2})$ with a beam waist $w_0$ of 23~$\mu$m and $\rho=\sqrt{X^2+Y^2}$. On any plane that contains the lattice axis, the lattice intensity distribution can be described by equal-intensity surfaces that resemble the shape of ellipses, as shown in Fig.~\ref{figs1}~(a). At the locations $(\rho=0, Z=k \frac{\lambda}{2})$, with integer $k$, the lattice laser intensity is maximal, $I=I_{\rm{max}}$.  For decreasing intensity, the equal-intensity ``ellipses'' become larger and increasingly eccentric, as shown in Fig.~\ref{figs1}~(a).

The atoms on the same equal-intensity surface have the same lower-transition detunings and Boltzmann factors; however, they experience different lattice-induced light shift for the 74S magic Rydberg state. The light shift of the 74S magic state is constant on cylindrical surfaces of constant $\rho$, {\sl{i.e.}} the 74S light shift is given by the lattice intensity $I(\rho,0)=I_{\rm{max}}\exp(\frac{-2\rho^2}{\omega_0^2})$. As a result, as shown in Fig.~\ref{figs1}(a), on any given constant-intensity ``ellipse'' with $I<I_{\rm{max}}$  there are two extreme cases: point A, at the major axis,  and point B, at the minor axis. Atoms at point A exhibit minimal light shift of the 74S state, and the signal strength due to atoms near A is maximal because the volume associated with the corresponding region is the largest. Atoms at point B experience maximal light shift of the 74S state, but the signal strength is small.

Recalling that the 5S and 5P$_{3/2}$ shifts are fixed on equal-intensity surfaces, the upper-transition detunings on equal-intensity surfaces vary over a frequency range $L$ given by the variation of the 74S light shift, as shown in Fig.~\ref{figs1}(b). Figure~\ref{figs1}(b) also shows that the range $L$ increases with decreasing value of the lower-transition detuning, $\Delta_{\rm{5S}} + \Delta_{\rm{5P}}$, which is seen as follows. The lower-transition light shift, $\Delta_{\rm{5S}} + \Delta_{\rm{5P}}$, has to approximately equal the 780-nm laser detuning $\Delta_{\rm{red}}$ for one of the light-shifted intermediate 5P$_{3/2}$ levels, so that level becomes resonantly populated. These resonances occur at certain intensities $I_{\rm{res}}$. The 74S light shift along the equal-intensity ``ellipse'' for $I_{\rm{res}}$ then varies between $-\frac{1}{4}\alpha_{\rm{74S}}\frac{2}{c\epsilon_0}I_{\rm{max}}$ at point B and $-\frac{1}{4}\alpha_{\rm{74S}}\frac{2}{c\epsilon_0}I_{\rm{res}}$ at point A. Hence, the frequency range $L$ associated with that instance of a doubly-resonant Rydberg excitation is
\begin{equation}
L = -\frac{1}{4}\alpha_{\rm{74S}}\frac{2}{c\epsilon_0}\big[I_{\rm{max}}-I_{\rm{res}}\big]
\end{equation}
This finding accords with the observation that for smaller lower-transition detunings $\Delta_{\rm {red}}$ the lattice spectral lines become wider. In addition, since the volumes associated with points along the equal-intensity surfaces increase in proportion with $\rho$, the signal increases when moving from points B to points A. As a result, with decreasing $\Delta_{\rm {red}}$ the lattice spectral lines become more asymmetric and triangular-shaped, as illustrated in Fig.~\ref{figs1}(c).

Most importantly, the spectral-line markers $(\Delta_{\rm {red}},\Delta_{\rm {blue}})$ used in the paper correspond to locations of the type A, where the lattice spectral lines peak. Hence, {\sl{all}} relevant light shifts (5S, 5P$_{3/2}$, and 74S) that correspond with the markers $(\Delta_{\rm {red}},\Delta_{\rm {blue}})$ are proportional to the same intensity $I(\rho,0)$. As a result, in our self-referencing method the intensity $I(\rho,0)$ drops out, allowing us to relate the 5P$_{3/2}$ polarizabilities to slopes in the modified a.c. Stark map. This proves one of the strongest features of our method, namely that the uncertainty of the calibration of lattice intensity against optical-lattice power injected into the cavity does not affect the uncertainty of the measured polarizability values.

\section{Statistical error}\label{waterfall}
\begin{figure*}
\begin{centering}
\includegraphics[width=7in]{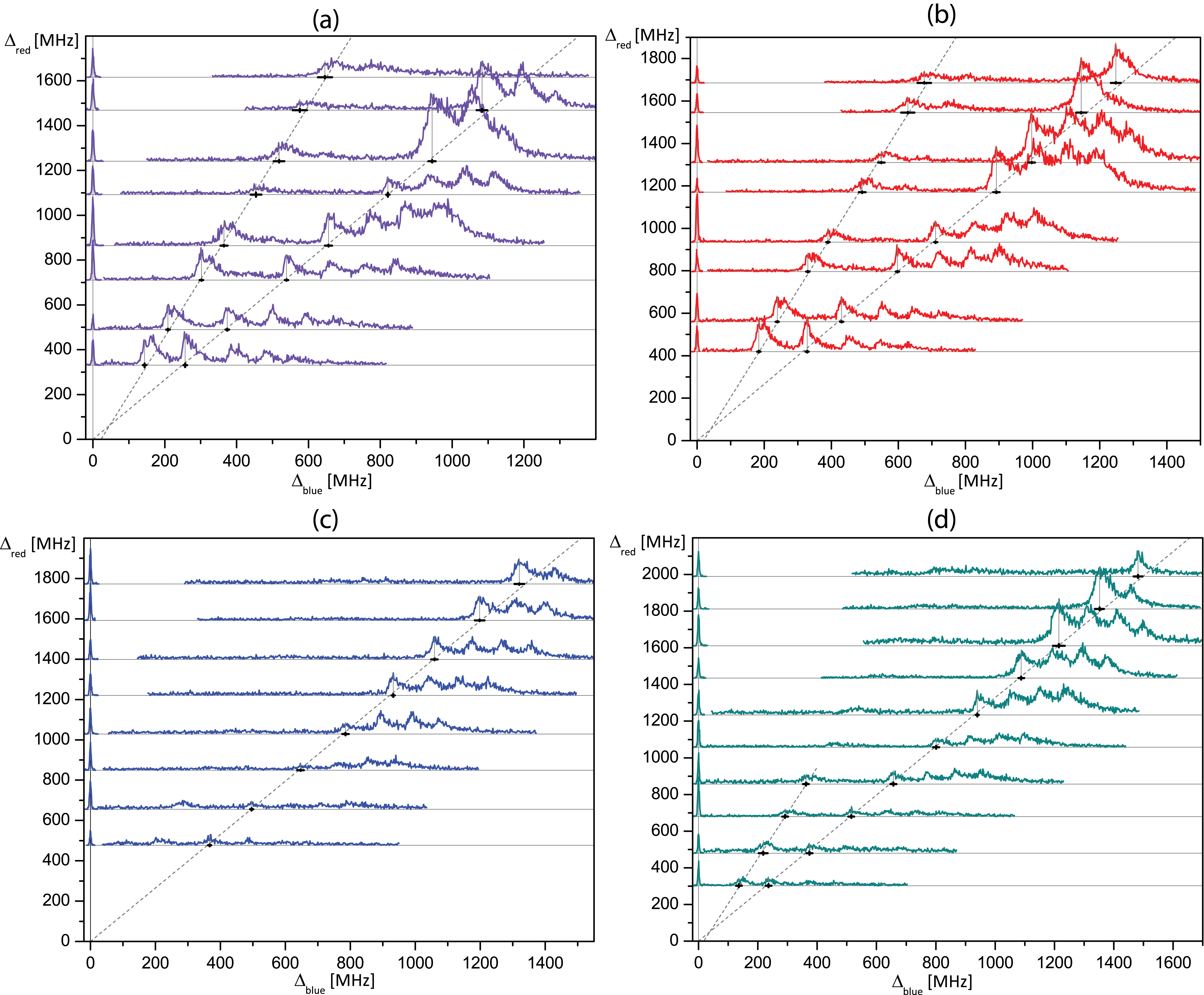}
\caption{(color online). Waterfall plots (modified a.c. Stark maps) of experimental spectra. 
The grey drop lines mark the $(\Delta_{\rm{blue}}, \Delta_{\rm{red}})$ coordinates of the peaks of the evaluated spectral lines.  The bold, short black crosses underneath the evaluated spectral lines indicate the uncertainty on the peak positions in both $\Delta_{\rm{blue}}$ and $\Delta_{\rm{red}}$.  The dashed grey lines show linear fits through the marked positions $(\Delta_{\rm{blue}}, \Delta_{\rm{red}})$. The transmitted lattice powers and the Rydberg excitation pulse durations are (a) 32~mW and 10~$\mu$s, (b) 36~mW and 10~$\mu$s, (c) 36~mW and 2$~\mu$s, (d) 40~mW and 10~$\mu$s.}
\label{figs2}
\end{centering}
\end{figure*}

We conduct the measurement at different lattice intensities and Rydberg-excitation durations (2~$\mu$s or 10~$\mu$s), as shown in Fig.~2 and Fig.~\ref{figs2}. With 10~$\mu$s Rydberg-excitation pulses, the lattice spectral lines tend to be saturated, but this does not shift the locations of the peak centers. Each data point $(\Delta_{\rm{blue}}, \Delta_{\rm{red}})$ in the modified a.c. Stark map has an individual uncertainty $(\delta(\Delta_{\rm{blue}}), \delta(\Delta_{\rm{red}}))$, which is due to the uncertainty associated with both excitation-laser frequency stabilization methods, the laser line widths, and the uncertainties in locating the peaks of the individual lines. We use OriginLab linear regression to determine the slopes $b$ of the highest energy levels in the $|m_j|=\frac{1}{2}$ groups and the $|m_j|=\frac{3}{2}$ groups. The fitting returns a fitting uncertainty $\delta b_1$. 

The effects of the reading uncertainties $(\delta(\Delta_{\rm{blue}}), \delta(\Delta_{\rm{red}}))$ is calculated as follows. Assuming there are $M$ data points for one energy level, with $x=\Delta_{\rm{blue}}$ and $y=\Delta_{\rm{red}}$. Each data point $(x=\Delta_{\rm{blue}}, y= \Delta_{\rm{red}})$ carries a Gaussian point spread function, with $\sigma_x=\delta(\Delta_{\rm{blue}})$ and $\sigma_y=\delta(\Delta_{\rm{red}})$. We generate $10^6$ sets of data points obtained with these point spread functions, and using random numbers. For each set, the slope $b$ of the best-fit linear 
function, $y=a+ b x$, is
\begin{eqnarray}
b &=&\frac{MXY-XY}{MX^2-XX}\\
X&=&\sum^M_{i=1}x_i \\
Y&=&\sum^M_{i=1}y_i \\
XY&=&\sum^M_{i=1}x_iy_i \\
X^2&=&\sum^M_{i=1}x_ix_i
\end{eqnarray}
The $10^6$ values of $b$ yield a distribution that has a standard deviation $\delta b_2$, which constitutes the uncertainty in $b$ due to the reading uncertainties $(\delta(\Delta_{\rm{blue}}), \delta(\Delta_{\rm{red}}))$.

The net statistical uncertainties $\delta b$ of the slopes are the quadrature sums of $\delta b_1$ and $\delta b_2$. In Table~\ref{tabs1} we show $b^i$ and $\delta b^i$ for the measurements performed, where $i$ is a counter for the measurement sets. We have five sets of scans, and obtain $N=4$ slopes for the highest level of the group $|m_j|=\frac{1}{2}$, and $N=5$ slopes for the group $|m_j|=\frac{3}{2}$. The averaged experimental values of the slopes and their uncertainties, $\bar{b}$ and $\bar{\delta b}$, are then calculated using weighted averages,
\begin{eqnarray}
\bar{b}          &=& \frac{\sum^N_{i=1}b^i (\delta b^i)^{-2}}{\sum^n_{i=1}(\delta b^i)^{-2}}\\
\bar{\delta b}^2 &=& \frac{1}{\sum^N_{i=1}(\delta b^i)^{-2}}
\end{eqnarray}

In the last step of the uncertainty analysis for the slopes, we add the relative uncertainties $\bar{\delta b}/\bar{b}$ and the scaling uncertainties, explained in the next section, in quadrature. This yields the net total uncertainties $\delta b_{\rm {net}}$ for both slopes. The final uncertainties for the polarizabilities are then calculated using standard uncertainty propagation in Eqs.~8 and 9.

\section{Scaling error}\label{error}
The lower (780~nm) and upper (480~nm) transition lasers of the Rydberg two-photon excitation are frequency-stabilized with the transmission spectra of confocal optical cavities. The lower transition utilizes a confocal cavity at 780~nm, whose free-spectral range (FSR) is 374.68 MHz with uncertainty of 0.50 MHz. This number is calibrated using the frequency reference of the well-known $^{85}$Rb transitions: $\rm{5S}_{1/2}\to\rm{5P}_{3/2}$ $\rm F=2\to F'=3$ and $\rm F=3\to F'=4$. The scaling error of $\Delta_{\rm{red}}$ is less than 0.2\%.

The upper transition 480~nm laser is frequency-doubled from a 960~nm laser, which is stabilized to a pressure-tuned confocal cavity at 960~nm. The FSR of the 960~nm confocal cavity is 489.37~MHz with uncertainty of 0.56 MHz. The calibration is referenced to a RF function generator (HP 8656B). The scans of the 480~nm laser are done by pressure-tuning the 960~nm FPI, which is driven by a step motor. In each of the scans, the raw data are recorded in terms of step numbers. We convert the step number into a MHz offset at 480~nm; the conversion is also referenced to a RF function generator. We found that the conversion factor from step number to MHz in 480~nm is not perfectly linear due to the nonlinearity of the pressure response to the step motor's position. During a single scan the nonlinearity is less than about 1.5\%. The resultant total scaling error of $\Delta_{\rm{blue}}$ is estimated to be less than 1.6\%.

\begin{table*}
\vspace{-12pt}
	\centering
	\caption{Slopes of the linear fits in the modified a.c. Stark maps from Fig.~2 and Fig.~\ref{figs2} and their uncertainties. The sample number is 5 for the slope of $|m_j|=\frac{3}{2}$, and 4 for the slope of $|m_j|=\frac{1}{2}$.}
		\begin{tabular}{lllllll}
		 \hline
		  figure~~~~~~~~~~~~& power (mW)~~~~& pulse ($\mu$s)~~~~& slope $|m_j|=\frac{3}{2}$~~~~& error~~~~~~~~~~~~& slope $|m_j|=\frac{1}{2}$~~~~& error \\ \hline
      Fig.2 & 20 & 2 & 2.45 & $6.95\times10^{-2}$ & 1.32 & $1.05\times10^{-2}$ \\
      Fig.S2 (a) & 32 & 10 & 2.58 & $9.46\times10^{-2}$ & 1.33 & $1.13\times10^{-2}$ \\
      Fig.S2 (b) & 36 & 10 & 2.53 & $8.15\times10^{-2}$ & 1.33 & $1.01\times10^{-2}$ \\
      Fig.S2 (c) & 36 & 2 & -- & -- & 1.32 & $8.14\times10^{-3}$ \\
       Fig.S2 (d) & 40 & 10 & 2.47 & $1.78\times10^{-1}$ & 1.33 & $8.17\times10^{-3}$ \\
       \hline \hline
		\end{tabular}
	\label{tabs1}
\end{table*}